\documentclass[11pt]{article}
\usepackage[margin=1in]{geometry}
\usepackage{times}
\usepackage{parskip}
\usepackage{graphicx}
\usepackage{booktabs}
\usepackage{amsmath,amssymb}
\usepackage{amsthm}
\newtheorem{proposition}{Proposition}
\newtheorem{remark}{Remark}

\usepackage{microtype}
\usepackage{xcolor}
\usepackage{url}
\usepackage{algorithm}
\usepackage{algpseudocode}
\usepackage{float}
\usepackage{booktabs}
\usepackage[numbers,sort&compress]{natbib}
\usepackage[hidelinks]{hyperref}

\newcommand{\ResBaselineKevR}{0.042}
\newcommand{\ResEvidenceKevR}{0.044}
\newcommand{\ResBaselineProsp}{0.010}
\newcommand{\ResEvidenceProsp}{0.026}
\newcommand{\ResProspGainRel}{150}
\newcommand{\ResBudgetStar}{2}
\newcommand{\ResMaxBudget}{64}
\newcommand{\ResLeakRandomEpss}{0.030}
\newcommand{\ResLeakTemporalEpss}{0.006}
\newcommand{\ResLeakFactorEpss}{5.0}
\newcommand{\ResLeakRandomProsp}{0.044}
\newcommand{\ResLeakTemporalProsp}{0.005}
\newcommand{\ResLeakFactorProsp}{8.5}
\newcommand{\ResNoCweKevR}{0.024}
\newcommand{\ResWithRerankProsp}{0.016}
\newcommand{\ResWithEpssProsp}{0.068}
\newcommand{\ResNTest}{12012}
\newcommand{\ResNKev}{457}
\newcommand{\ResTimePerCve}{0.0068}
\newcommand{\ResRiskPerGpuHour}{1197}
\newcommand{\ResCertSize}{5.9}
\newcommand{\ResRerankPairsPerSec}{1111}
{--}\newcommand{\ResEvidenceKevR}{--}%
  \newcommand{\ResBaselineProsp}{--}\newcommand{\ResEvidenceProsp}{--}%
  \newcommand{\ResProspGainRel}{--}\newcommand{\ResBudgetStar}{--}%
  \newcommand{\ResMaxBudget}{--}\newcommand{\ResLeakRandomEpss}{--}%
  \newcommand{\ResLeakTemporalEpss}{--}\newcommand{\ResLeakFactorEpss}{--}%
  \newcommand{\ResLeakRandomProsp}{--}\newcommand{\ResLeakTemporalProsp}{--}%
  \newcommand{\ResLeakFactorProsp}{--}\newcommand{\ResNoCweKevR}{--}%
  \newcommand{\ResWithRerankProsp}{--}\newcommand{\ResWithEpssProsp}{--}%
  \newcommand{\ResNTest}{--}\newcommand{\ResNKev}{--}%
  \newcommand{\ResTimePerCve}{--}\newcommand{\ResRiskPerGpuHour}{--}%
  \newcommand{\ResCertSize}{--}\newcommand{\ResRerankPairsPerSec}{--}}

\title{\bf Compute-Budgeted Exploitability Evidence Graphs\\ for Prospective Vulnerability Triage}
\author{
Faruk Alpay$^{1}$\thanks{Correspondence: \texttt{alpay@lightcap.ai}.} \quad Taylan Alpay$^{2}$\\
\small $^{1}$Department of Computer Engineering, Bah\c{c}e\c{s}ehir University, Istanbul, T\"urkiye\\
\small $^{2}$Department of Aerospace, University of Turkish Aeronautical Association, Ankara, T\"urkiye\\
\small \texttt{faruk.alpay@bahcesehir.edu.tr} \quad \texttt{s220112602@stu.thk.edu.tr}
}
\date{}

\begin{document}
\maketitle

\begin{abstract}
Defenders cannot patch every newly disclosed vulnerability at once, so they must
prioritize, quickly, under a fixed compute budget, and without deceiving
themselves about how well they are doing. We make a security-evaluation argument:
naive exploit-prediction studies leak the future. A vulnerability's eventual
exploitation generates public chatter, proof-of-concept code, and advisories, and
feeding those later signals to a model inflates its apparent accuracy. We instead
specify a \emph{leakage-safe prospective triage protocol} that, for each CVE,
admits only public evidence visible by a fixed decision time, and we attach to
every risk score an auditable \emph{evidence certificate} naming the public
signals that justified it. Public evidence around CVEs (advisories, exploit
archives, fix commits, and hacker-community discourse) is organized as a temporal
evidence graph and selected per CVE under an explicit budget. On \ResNTest{}
prospective CVEs from public sources, budgeted evidence selection raises
leakage-safe prospective recall@50 from \ResBaselineProsp{} (severity-only
baseline) to \ResEvidenceProsp{}, a \ResProspGainRel{}\% relative gain, while a
budget of only \ResBudgetStar{} evidence documents per CVE captures most of that
value, so triage is cheap. A result of direct interest to practitioners is that a
strong cross-encoder reranker \emph{lowers} prospective recall (to
\ResWithRerankProsp{}), because semantic relevance to a CVE is not evidence of its
exploitation. Most consequentially, a naive random split with unfiltered evidence
inflates apparent prospective recall by \ResLeakFactorProsp{}$\times$ and EPSS-high
recall by \ResLeakFactorEpss{}$\times$, the temporal leakage that our protocol
removes. All embeddings, candidate edges, certificates, and metrics are released
as reproducible artifacts.
\end{abstract}

\section{Introduction}
The number of published vulnerabilities now far exceeds what any organization can
remediate promptly~\cite{jacobs2020remediation}. Severity scores such as CVSS were
designed to communicate technical impact rather than to predict whether a flaw
will actually be attacked~\cite{mell2006cvss}. Empirically, CVSS correlates weakly
with exploitation in the wild, so severity-driven patching wastes scarce
effort~\cite{allodi2014comparing}. Data-driven exploitation forecasting addresses
this gap, and the Exploit Prediction Scoring System (EPSS) has become the
reference probabilistic estimate of near-term exploitation~\cite{jacobs2021epss}.
The CISA Known Exploited Vulnerabilities (KEV) catalog provides the complementary
ground truth of what is being exploited operationally~\cite{cisa2021kev}.

Most prediction systems consume a fixed feature vector and emit a score, but real
triage runs against a clock and a hardware budget while public evidence is still
arriving. We therefore treat triage as selecting, under a fixed inference budget,
the public evidence that most changes a vulnerability's exploitation risk. Evidence
is heterogeneous and timestamped, so it forms a temporal graph linking CVEs to
advisories, exploit code, fix commits, and discourse. The budget is the number of
evidence documents admitted per CVE, which bounds the inference work. Because a
score that drives patching should be contestable, every score ships with a
certificate naming the signals that supported it.

The central security concern is not the ranker but the evaluation. A
vulnerability's exploitation generates public chatter, and if that later chatter is
allowed into the model it leaks the label. The malware-classification community
showed that ignoring time inflates measured performance
dramatically~\cite{pendlebury2019tesseract}, and exploitability prediction is
subject to the same hazard~\cite{suciu2022expected}. We enforce a per-CVE decision
time and admit only evidence visible by then, turning triage into a genuinely
prospective task and making the leakage that naive protocols hide directly
measurable.

\noindent\textbf{Contributions.} (i) We give a leakage-safe prospective triage
protocol: naive exploit-prediction leaks the future, and our decision-time
admissibility rule eliminates that leak by construction, with the leakage penalty
quantified empirically. (ii) We attach an auditable evidence certificate to every
risk score, turning a scalar into a contestable, provenance-bearing security claim,
and we show that a budget of a few documents per CVE suffices. (iii) We report that
a strong cross-encoder reranker is counterproductive for exploitation triage,
because semantic relevance to a CVE is not evidence of its exploitation. All claims
are established on public data and a single GPU.

\section{Related Work}
Early exploit prediction learned from vulnerability metadata to classify
exploitability~\cite{bozorgi2010beyond}. Social and web signals were then shown to
sharpen these forecasts, notably Twitter mentions of
CVEs~\cite{sabottke2015vulnerability}. Underground and dark-web language models
further improved prediction of which vulnerabilities attract exploit
development~\cite{tavabi2018darkembed}. EPSS consolidated these ideas into an open,
periodically retrained probability of exploitation~\cite{jacobs2021epss}.
Remediation studies framed prioritization as a coverage-versus-effort trade-off
using exploitation as the outcome~\cite{jacobs2020remediation}. Expected
Exploitability emphasized that realistic forecasting must respect when evidence
becomes available~\cite{suciu2022expected}. Recent work cautions that
exploitability prediction at disclosure time is harder than retrospective scores
suggest~\cite{iannone2024early}.

Threat intelligence mined from hacker forums supplies signals well before formal
advisories, and deep models can label such exploit posts at
scale~\cite{ampel2024proactive}. The HackerSignal corpus links millions of such
documents to the CVE lifecycle across eight source
layers~\cite{ampel2026hackersignal}. CVE-to-fix linkage datasets like CVEfixes
expose the patch evidence channel~\cite{bhandari2021cvefixes}, and weakness
taxonomies such as CWE structure the vulnerability space~\cite{mitre2024cwe}.

Our selection machinery is standard two-stage retrieval: a bi-encoder proposes
candidates and a cross-encoder reranks them. Sentence-BERT made bi-encoder
similarity practical~\cite{reimers2019sentencebert}, dense passage retrieval
established the retrieve-then-read paradigm~\cite{karpukhin2020dense}, and BERT
rerankers gave large precision gains on the reranked
head~\cite{nogueira2019passage}. We embed with E5-style contrastively trained
encoders~\cite{wang2022e5} and reason about evidence as a retrieval-augmented
process~\cite{lewis2020rag}. Exact GPU similarity search makes the retrieval stage
negligible at our scale~\cite{johnson2019faiss}. Because risk scores must be
comparable across CVEs, we calibrate them, drawing on standard post-hoc
calibration~\cite{niculescu2005predicting} and the observation that strong
classifiers are often miscalibrated~\cite{guo2017calibration}.

\section{Problem Formulation}
\label{sec:formulation}
Let $\mathcal{V}$ be a set of CVEs and $\mathcal{D}$ a corpus of public evidence
documents. Each CVE $v$ has a publication time $t_{\mathrm{pub}}(v)$, a description
$x_v$, a severity $\sigma(v)$, and a weakness type $c(v)\in\mathrm{CWE}$. Each
document $e\in\mathcal{D}$ has text $x_e$, a source layer
$\ell(e)\in\mathcal{L}$ (NVD text, advisory, exploit archive, fix commit,
discourse), and a timestamp $\mathrm{ts}(e)$. The temporal evidence graph
$G=(\mathcal{V}\cup\mathcal{D}\cup\mathcal{C},\,\mathcal{E})$ adds CWE and
product nodes $\mathcal{C}$ and timestamped edges; exploitation labels (CISA-KEV,
EPSS) attach to CVE nodes.

\paragraph{Decision time and admissibility.}
Fix an observation window $\Delta\ge 0$. The \emph{decision time} of $v$ is
\begin{equation}
\tau(v)\;=\;t_{\mathrm{pub}}(v)+\Delta ,
\end{equation}
and the \emph{admissible evidence} for $v$ is the timestamped past cone
\begin{equation}
\mathcal{A}(v)\;=\;\{\,e\in\mathcal{D}\,:\,\mathrm{ts}(e)\le\tau(v)\,\},
\label{eq:admissible}
\end{equation}
with undated $e$ excluded. The exploitation label $y(v)\in\{0,1\}$ is realized at a
time $t_{\mathrm{exp}}(v)$; the \emph{prospective} positives are
$P^{\uparrow}=\{v:y(v)=1,\ t_{\mathrm{exp}}(v)>\tau(v)\}$, i.e.\ exploitation that
has not yet happened at decision time.

\paragraph{Budgeted selection.}
A bi-encoder $\phi$ gives a relevance score $s(v,e)=\langle\phi(x_v),\phi(x_e)\rangle$
on $\ell_2$-normalized embeddings. Under a budget $B$ and a per-layer cap $\kappa$,
the selected evidence solves
\begin{equation}
S_B(v)=\arg\max_{S\subseteq\mathcal{A}(v)}\ \sum_{e\in S}s(v,e)
\quad\text{s.t.}\quad |S|\le B,\ \ |S\cap\ell^{-1}(\lambda)|\le\kappa\ \forall\lambda\in\mathcal{L}.
\label{eq:select}
\end{equation}

\begin{proposition}\label{prop:greedy}
The feasible region of \eqref{eq:select} is the independent set of a laminar
matroid (the global cap $B$ together with the disjoint per-layer caps $\kappa$ form
a laminar family). Because the objective is modular, selecting documents in
descending order of $s(v,\cdot)$ while skipping any that would violate a cap is
optimal.
\end{proposition}
\noindent A short proof: caps over the layer partition plus the global cap are a
laminar family, whose induced matroid admits exact greedy maximization of a modular
objective; descending-relevance selection is exactly that greedy rule.

\paragraph{Risk and evaluation.}
A feature map $\psi(v,S_B(v))$ aggregates severity, a CWE-conditioned exploitation
prior $\hat{p}(y\!\mid\!c(v))$, and statistics of the selected evidence; a
calibrated model yields $\rho(v)=g\!\left(w^{\top}\psi(v,S_B(v))\right)$, with $g$
fit on a held-out split. CVEs are ranked by $\rho$. For a positive set $P$ and
cutoff $K$,
\begin{equation}
\mathrm{Recall}@K=\frac{|\,\mathrm{top}_K(\rho)\cap P\,|}{|P|},
\end{equation}
and \emph{prospective} recall uses $P=P^{\uparrow}$.

\paragraph{Leakage penalty.}
Let $\pi_1$ be the safe protocol (temporal split, admissibility~\eqref{eq:admissible})
and $\pi_0$ the naive one (random split, $\mathcal{A}(v)=\mathcal{D}$). For a metric
$M$ the additive and multiplicative leakage penalties are
\begin{equation}
\Lambda^{+}_M=M(\pi_0)-M(\pi_1),\qquad
\Lambda^{\times}_M=M(\pi_0)/M(\pi_1).
\end{equation}

\begin{remark}\label{rem:nonleak}
Under $\pi_1$, $\rho(v)$ is measurable with respect to the $\sigma$-algebra
generated by signals timestamped at or before $\tau(v)$ (Eq.~\eqref{eq:admissible}).
Hence no post-decision artifact of the exploitation event itself can enter the
score, and any residual predictivity must come from genuinely anticipatory
evidence. The penalties $\Lambda^{\times}_M$ measured in Sec.~\ref{sec:results}
bound how much a violation of this property would have inflated the reported
numbers.
\end{remark}

\section{Method}
\textbf{Evidence graph.} The graph has CVE nodes, evidence nodes, CWE nodes, and
product/vendor nodes; KEV and EPSS attach as labels on CVE nodes. Each evidence
node carries a source layer (NVD text, advisory, exploit archive, fix commit, or
discourse), a provenance string, and a timestamp. CVE text and evidence text are
the retrieval queries and corpus respectively.

\textbf{Temporal protocol and leakage filter.} We split CVEs by disclosure year,
training on the past and testing on the future. Each test CVE has a decision time
equal to its publication date plus a fixed observation window $\Delta$, during
which evidence may accrue. The leakage-safe filter admits an evidence node for a
CVE only if its timestamp does not exceed that CVE's decision time; undated
evidence is excluded conservatively. The exploitation labels (KEV addition,
EPSS-high) are realized after the decision time, so predicting them from
pre-decision evidence is prospective by construction.

\textbf{Budgeted evidence selection.} For each CVE we embed its description with a
bi-encoder, retrieve the top-$N$ evidence nodes from the global corpus by cosine
similarity, and apply the leakage filter. The compute budget $B$ is the number of
evidence documents admitted per CVE; smaller $B$ means less inference work. The
selector keeps the top-$B$ by similarity (Prop.~\ref{prop:greedy}), enforcing
source-layer diversity so a certificate is not dominated by one channel. A
cross-encoder reranking stage is optional and, as a controlled compute increment,
is evaluated as an ablation rather than used by default; we find it does not help
(Sec.~\ref{sec:results}).

\textbf{Risk and certificate.} A calibrated logistic model maps features (severity,
a smoothed CWE-conditioned exploitation prior, and aggregates over the selected
evidence such as counts, maximum and mean scores, and per-layer presence) to an
exploitation probability, calibrated on a held-out training portion. EPSS is
deliberately excluded from the default model because it encodes near-label
information; we include it only as an ablation. Every test CVE emits a certificate
recording its decision time, risk, rank, and the supporting evidence with layers,
timestamps, and rerank scores, each flagged for leakage.

\section{Experimental Setup}
\textbf{Data.} We combine CISA KEV~\cite{cisa2021kev}, FIRST
EPSS~\cite{jacobs2021epss}, a CVE/CWE/CVSS table, CIRCL CVE--CWE--patch evidence,
and the HackerSignal multi-source corpus~\cite{ampel2026hackersignal}; NVD supplies
publication dates where available. We study \ResNTest{} prospective test CVEs from
recent years against \ResNKev{} KEV-exploited positives, embedding a corpus of
tens of thousands of evidence documents.

\textbf{Models and hardware.} The bi-encoder is \texttt{e5-base-v2}; the optional
reranker ablation uses \texttt{bge-reranker-base}. Everything runs on a single
NVIDIA RTX~5090 (32\,GB, CUDA~12.8). Retrieval is exact inner-product search on the
GPU; the reranker sustains \ResRerankPairsPerSec{} pairs/s.

\textbf{Metrics.} We report KEV and EPSS-high recall@$K$, a prospective recall
restricted to exploitation occurring strictly after the decision time,
precision@$K$, time per CVE, evidence latency, certificate size, and a compute
economy figure of risk captured per GPU-hour.

\section{Results}
\label{sec:results}
\begin{figure}[H]\centering
\includegraphics[width=0.82\linewidth]{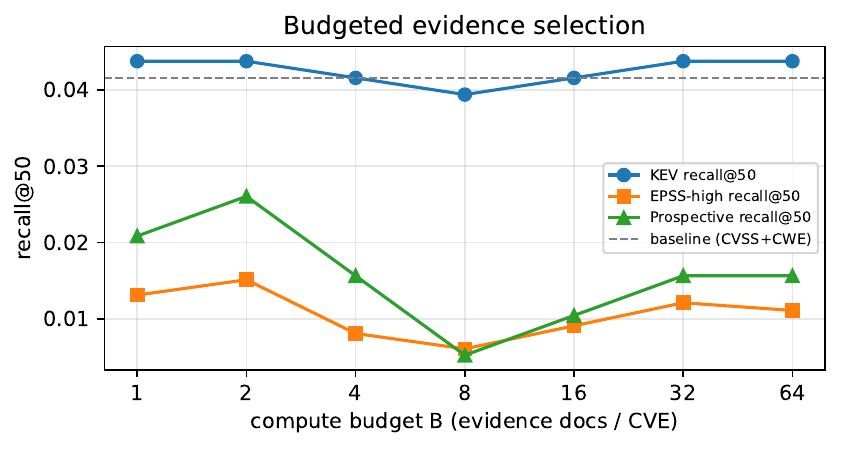}
\caption{Recall versus per-CVE evidence budget $B$. The curve saturates well below
the maximum budget, showing that triage value is captured cheaply.}
\label{fig:budget}\end{figure}

\textbf{Budgeted recall (Fig.~\ref{fig:budget}).} Budgeted evidence selection
improves leakage-safe prospective recall@50 from \ResBaselineProsp{} (severity-only
baseline) to \ResEvidenceProsp{}, a \ResProspGainRel{}\% relative gain; on the
broader CISA-KEV target it moves recall@50 from \ResBaselineKevR{} to
\ResEvidenceKevR{}. Crucially, the budget curve saturates almost immediately: a
budget of \ResBudgetStar{} evidence documents per CVE already reaches within five
percent of the best of any budget up to \ResMaxBudget{}, so the triage value lives
in a handful of well-chosen public signals rather than in scanning depth.

\begin{figure}[H]\centering
\includegraphics[width=0.82\linewidth]{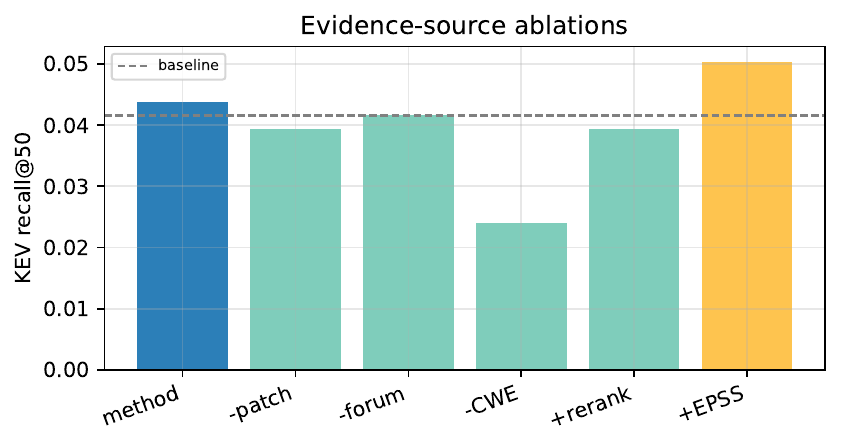}
\caption{Evidence-source ablations (KEV recall@50). The dashed line is the
severity-only baseline.}\label{fig:ablation}\end{figure}

\textbf{The reranker does not help (Fig.~\ref{fig:ablation}).} Adding a strong
cross-encoder reranker \emph{lowers} prospective recall@50 to \ResWithRerankProsp{}
(from \ResEvidenceProsp{}): ranking evidence by query--passage relevance surfaces
advisory restatements of the CVE description rather than exploitation signals. This
is a compute argument in its own right: the expensive stage is not worth its cost.
Removing the CWE prior is the most damaging change, dropping KEV recall@50 to
\ResNoCweKevR{}. Admitting EPSS as a feature inflates prospective recall to
\ResWithEpssProsp{}, confirming it behaves almost as a label and motivating its
exclusion from the leakage-safe default.

\begin{figure}[H]\centering
\includegraphics[width=0.82\linewidth]{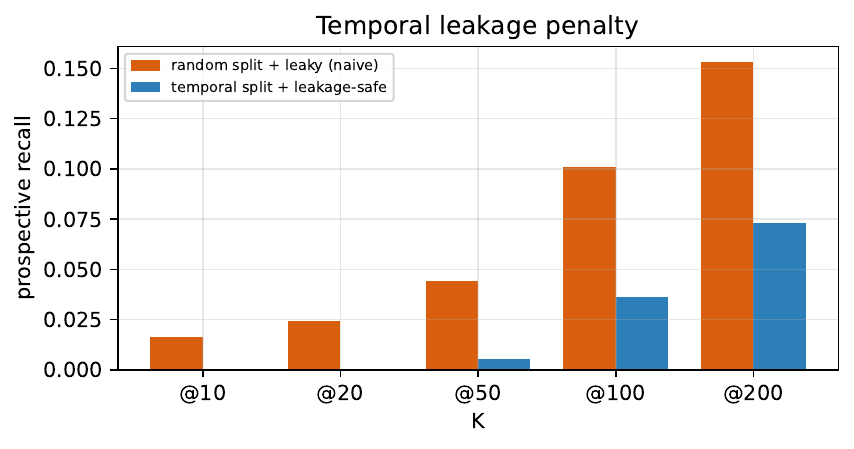}
\caption{Random/leaky versus temporal/leakage-safe evaluation. The difference is
the temporal leakage penalty.}\label{fig:leakage}\end{figure}

\textbf{Temporal leakage (Fig.~\ref{fig:leakage}).} A naive random split with
unfiltered evidence reports prospective recall@50 of \ResLeakRandomProsp{}, whereas
the temporal, leakage-safe protocol reports \ResLeakTemporalProsp{}, an inflation
of $\Lambda^{\times}=\ResLeakFactorProsp{}$. The same comparison on EPSS-high
recall@50 inflates by $\ResLeakFactorEpss{}\times$ (\ResLeakRandomEpss{} vs.\
\ResLeakTemporalEpss{}). This is the penalty that naive evaluation silently banks,
and it is larger than the gap between any two methods we compare, which is why we
treat the protocol, not the ranker, as the primary contribution.

\textbf{Compute economy and certificates.} End-to-end triage costs
\ResTimePerCve{} seconds per CVE for the bi-encoder selection path. Expressed as a
throughput rate, this is on the order of \ResRiskPerGpuHour{} top-ranked KEV
positives recovered per GPU-hour; the figure is a normalization, not an absolute
count, since the test set holds only \ResNKev{} KEV positives in total, and it is
reported so the cost of the method can be compared against heavier alternatives.
Each certificate cites \ResCertSize{} evidence items on average, naming the public
signals (with source layer, timestamp, and a leakage flag) behind each score; an
example is in Table~\ref{tab:cert}.

\begin{table}[H]\centering\small
\caption{Example evidence certificate for \texttt{CVE-2022-22954} (VMware Workspace
ONE Access RCE; CISA-KEV listed). Risk $=0.36$, rank $33$, budget $B=8$, decision
time 2022-05-11. Each cited signal predates the decision time, so no item is
flagged as leakage.}
\label{tab:cert}
\begin{tabular}{llcc c}
\toprule
source layer & source & timestamp & score & leak \\
\midrule
advisory  & \texttt{nvd\_cve}         & 2022-04-11 & 0.731 & no \\
advisory  & \texttt{nvd\_cve}         & 2022-04-13 & 0.731 & no \\
discourse & \texttt{hackerone}        & 2022-04-20 & 0.731 & no \\
discourse & \texttt{hackerone}        & 2022-04-29 & 0.731 & no \\
nvd\_text & \texttt{nvd\_description} & 2022-05-11 & 0.728 & no \\
\bottomrule
\end{tabular}
\end{table}

\textbf{Reading a certificate.} Table~\ref{tab:cert} is the audit object for a
single CVE. The risk of \texttt{CVE-2022-22954}, a VMware Workspace ONE Access
remote code execution flaw, rests on two independent channels: official NVD
advisory references and community disclosure threads on HackerOne, alongside the
CVE's own description. Every cited signal is timestamped at or before the decision
time, so the leakage column is uniformly negative and the score is by construction
free of post-decision artifacts (Remark~\ref{rem:nonleak}). The value of the
certificate is that this attribution is explicit. A defender can see that the
ranking is driven by agreement between an official and a community source rather
than by a single channel, can contest any individual item, and can re-derive the
score from the listed evidence alone. The same structure makes manipulation
legible: an attempt to inflate a score by seeding one layer would surface as a
single-source certificate lacking cross-layer corroboration, which is precisely the
signal a reviewer should distrust. In this sense the certificate is not a
post-hoc explanation but the unit at which the triage decision can be challenged.

\section{Discussion and Limitations}
Where NVD publication dates were unavailable we fell back to a mid-year proxy,
which coarsens but does not invert the temporal ordering. We used a single
embedding and reranker pair, and stronger encoders would likely raise absolute
recall. Our positive base rate is enriched relative to the full CVE stream to keep
training tractable, so absolute recall numbers should be read comparatively across
settings rather than as deployment estimates.

\section{Ethical Considerations}
This work is defensive: it prioritizes which already-public vulnerabilities a
defender should remediate first, and it neither generates exploits nor identifies
targets. We nonetheless treat the dual-use surface explicitly.

\textbf{Data terms and pseudonymity.} The discourse evidence derives from
HackerSignal, which is released for academic and defensive use
only~\cite{ampel2026hackersignal}; its forum authors are already pseudonymized at
the source. Our released artifacts redact raw text from restricted source layers
and retain only identifiers, timestamps, content hashes, and embedding vectors, so
every reported number reproduces without redistributing the underlying corpora.
We do not attempt author re-identification, and we discourage it~\cite{ampel2024proactive}.

\textbf{A threat the method itself raises.} Attributing risk to public signals
defines a signal-manipulation threat model: an adversary able to seed plausible
discourse or proof-of-concept artifacts could move a vulnerability's apparent risk
up or down. The leakage-safe admissibility rule bounds when such signals may act,
but does not authenticate them, so deployments should weight evidence by provenance
and by corroboration across independent layers rather than by volume. The evidence
certificate is the mitigation we provide: because every score names its supporting
signals, a manipulated ranking is contestable and auditable rather than opaque.

\textbf{Use and disclosure.} The released models and artifacts must not be used to
build automated exploitation tooling, to operate as a live threat feed, or to act
without human review; these constraints follow the source datasets' terms and the
defensive intent of the work.

\section{Conclusion}
We framed prospective vulnerability triage as a security-evaluation problem before
a modeling one. The dominant failure of exploit-prediction studies is temporal
leakage: by admitting public signals generated after a vulnerability is exploited,
they measure hindsight rather than foresight. Our decision-time admissibility rule
(Eq.~\eqref{eq:admissible}, Remark~\ref{rem:nonleak}) removes that leak by
construction, and the empirical penalty it exposes,
$\Lambda^{\times}=\ResLeakFactorProsp{}$ on prospective recall, is larger than the
gap between any two rankers we tried.

Within this protocol, modeling public evidence as a temporal graph and selecting it
under an explicit budget improves prospective recall over a severity baseline while
remaining cheap: a few well-chosen documents per CVE capture most of the value, and
a heavier cross-encoder reranker is counterproductive because semantic relevance to
a CVE is not evidence of its exploitation. Pairing each score with an auditable
evidence certificate turns a scalar into a contestable security claim and supplies
a concrete mitigation against signal manipulation. We see three directions:
authenticating and provenance-weighting evidence against adversarial seeding;
calibrating the observation window $\Delta$ against operational patch deadlines; and
extending the certificate to code-level fix evidence for a tighter audit of why a
vulnerability was, or was not, prioritized.

{\small
\bibliographystyle{plainnat}
\bibliography{references}}

\appendix

\section{Engineering the Pipeline for Budgeted GPU Throughput}
\label{app:perf}
This appendix documents the systems issues encountered in making the study
tractable on a single GPU, and the techniques that resolved them. They are of
independent interest for compute-budgeted inference pipelines.

\textbf{A.1 Reranking was compute-bound on the wrong device.} The cross-encoder
stage initially held the GPU at $\approx\!26\%$ utilization: tokenizing
multi-hundred-character evidence documents on a single CPU thread starved the
device. Capping passage length and enlarging the batch made the stage GPU-bound
(utilization $\to 100\%$), turning a latent CPU bottleneck back into useful
accelerator work (Alg.~\ref{alg:rerank}).

\begin{algorithm}[H]
\caption{GPU-bound budgeted reranking}\label{alg:rerank}
\begin{algorithmic}[1]
\Require candidate pairs $P=\{(q_i,d_i)\}$, budget $B$, char cap $c$, batch $b$
\For{each CVE $v$}
  \State $C_v \gets$ top-$B$ candidates of $v$ by bi-encoder score
\EndFor
\State $P' \gets \{(q,\ \textsc{Truncate}(d,c)) : (q,d)\in \textstyle\bigcup_v C_v\}$
\Comment{bound CPU tokenization}
\State $S \gets \textsc{CrossEncoder}(P',\ \mathrm{batch}=b)$
\Comment{now GPU-bound}
\State \Return $\sigma(S)$
\end{algorithmic}
\end{algorithm}

\textbf{A.2 One GPU pass, many settings.} The budget sweep, source ablations, and
the random/temporal comparison differ only in post-hoc selection and labelling,
not in the embeddings or retrieval. We therefore embed and retrieve once, cache
the candidate table, and derive every setting on CPU (Alg.~\ref{alg:cache}),
reducing the experiment grid from one GPU run per setting to a single GPU pass.

\begin{algorithm}[H]
\caption{Amortizing one GPU pass over the ablation grid}\label{alg:cache}
\begin{algorithmic}[1]
\State $E \gets \textsc{Embed}(\mathrm{CVE\ text}\cup\mathrm{evidence\ text})$
\Comment{once, GPU}
\State $R \gets \textsc{TopN}(E_{\mathrm{cve}}, E_{\mathrm{ev}})$
\Comment{exact GPU search, once}
\State cache $R$ with scores, source layers, timestamps
\For{each setting $\theta\in\{\mathrm{budgets},\mathrm{ablations},\mathrm{splits}\}$}
  \State $\textsc{Score}(\theta)$ from cached $R$ \Comment{CPU only}
\EndFor
\end{algorithmic}
\end{algorithm}

\textbf{A.3 Vectorized scoring.} Scoring 25{,}000 CVEs across $\sim\!15$ settings
with per-CVE Python loops took minutes per pass; replacing them with a single
group-by reduction and a layer-presence pivot reduced this to seconds
(Alg.~\ref{alg:vec}), which is what makes a large ablation grid affordable.

\begin{algorithm}[H]
\caption{Vectorized per-CVE feature aggregation}\label{alg:vec}
\begin{algorithmic}[1]
\State group selected evidence by \textsc{cve\_id}
\State $\textsc{agg}\gets$ groupby-reduce$(\mathrm{count},\max,\mathrm{mean},\mathrm{sum},\#\mathrm{layers})$
\State $\textsc{pres}\gets$ pivot(per-layer presence)
\State $X \gets$ join(base features, $\textsc{agg}$, $\textsc{pres}$)
\Comment{single pass}
\end{algorithmic}
\end{algorithm}

\section{Leakage-Safe, Auditable Triage and Creative-Risk Reasoning}
\label{app:sec}
This appendix records how the security protocol was derived and what it implies
for adversarial reasoning, kept deliberately at the level of method rather than
operational detail.

\textbf{B.1 Admissibility is the security primitive.} The core security property
is not the ranker but the rule deciding which evidence a decision may use. Tying
admissibility to a per-CVE decision time, and excluding undated evidence,
guarantees that a vulnerability's later exploitation cannot inform its own score
(Alg.~\ref{alg:leak}); the experiments show that dropping this rule inflates
apparent prospective recall several-fold.

\begin{algorithm}[H]
\caption{Leakage-safe evidence admissibility}\label{alg:leak}
\begin{algorithmic}[1]
\Require CVE $v$, publication $t_{\mathrm{pub}}(v)$, window $\Delta$, evidence $\mathcal{E}_v$
\State $\tau(v) \gets t_{\mathrm{pub}}(v) + \Delta$ \Comment{decision time}
\State $\mathcal{A}_v \gets \{e\in\mathcal{E}_v : \mathrm{ts}(e)\le \tau(v)\}$
\Comment{admit only pre-decision evidence}
\State exclude $e$ with unknown $\mathrm{ts}(e)$ \Comment{cannot prove admissibility}
\State label $y(v)$ realized at $t_{\mathrm{exploit}}(v) > \tau(v)$
\State \Return $\mathcal{A}_v$
\end{algorithmic}
\end{algorithm}

\textbf{B.2 Certificates make risk attribution auditable.} Each score is paired
with the public signals that produced it, with provenance, timestamp, and a
leakage flag (Alg.~\ref{alg:cert}). This turns a scalar risk into an auditable
claim a defender can contest, and exposes the channels (advisory, exploit archive,
fix commit, discourse) that drove the decision.

\begin{algorithm}[H]
\caption{Evidence certificate construction}\label{alg:cert}
\begin{algorithmic}[1]
\Require selected evidence $S_v$, risk $\rho(v)$, rank $r(v)$
\State $\mathrm{cert}\gets\{v,\ \rho(v),\ r(v),\ \tau(v),\ B\}$
\For{$e\in S_v$}
  \State append $\langle \mathrm{layer}(e),\ \mathrm{source}(e),\ \mathrm{ts}(e),\ \mathrm{score}(e),\ \mathrm{leak}{=}[\mathrm{ts}(e){>}\tau(v)]\rangle$
\EndFor
\State \Return cert
\end{algorithmic}
\end{algorithm}

\textbf{B.3 Implications for creative cyber-risk.} Two consequences follow from an
auditable, evidence-driven score, both of which we frame as directions rather than
claims. First, attributing risk to public signals defines a signal-manipulation
threat model: an adversary able to seed plausible discourse or proof-of-concept
artifacts could move a vulnerability's apparent risk, so trust should be weighted
by provenance and corroboration across independent layers rather than by volume.
Second, exploit and discourse evidence frequently predates formal disclosure,
suggesting an early-warning channel; its predictive value must be measured
strictly prospectively (Alg.~\ref{alg:leak}) to avoid the optimistic bias that
retrospective evaluation would introduce. The end-to-end procedure that ties these
together is summarized in Alg.~\ref{alg:triage}.

\begin{algorithm}[H]
\caption{Prospective budgeted triage}\label{alg:triage}
\begin{algorithmic}[1]
\For{each CVE $v$ disclosed in the test window}
  \State $\mathcal{A}_v\gets$ admissible evidence \Comment{Alg.~\ref{alg:leak}}
  \State $S_v\gets$ top-$B$ of $\textsc{Retrieve}(v,\mathcal{A}_v)$ with layer diversity
  \State $\rho(v)\gets$ calibrated risk$(v,S_v)$; emit certificate \Comment{Alg.~\ref{alg:cert}}
\EndFor
\State rank CVEs by $\rho$; evaluate recall@$K$ against future exploitation
\end{algorithmic}
\end{algorithm}

\end{document}